\documentclass[preprintnumbers, floatfix,letterpaper,aps,prd,epsfig,nofootinbib,
longbibliography,
twocolumn
]{revtex4-1}
\usepackage{bm,graphicx,dcolumn,epstopdf,epsf, latexsym,mathbbol,amssymb,amsmath, color, slashed, mathrsfs,mathcomp,simplewick}
\usepackage[center]{subfigure}
\usepackage{multirow}
\usepackage{makecell}
\usepackage{ytableau}
\usepackage{booktabs}
\usepackage{lipsum}
\usepackage[colorlinks,linkcolor=blue,citecolor=blue,urlcolor=blue]{hyperref}

\begin{document}
\allowdisplaybreaks
 \newcommand{\bq}{\begin{equation}}
 \newcommand{\eq}{\end{equation}}
 \newcommand{\bqn}{\begin{eqnarray}}
 \newcommand{\eqn}{\end{eqnarray}}
 \newcommand{\nb}{\nonumber}
 \newcommand{\lb}{\label}
 \newcommand{\f}{\frac}
 \newcommand{\p}{\partial}
 \newcommand{\bl}{\boldsymbol}
\newcommand{\PRL}{Phys. Rev. Lett.}
\newcommand{\PLB}{Phys. Lett. B}
\newcommand{\PRD}{Phys. Rev. D}
\newcommand{\CQG}{Class. Quantum Grav.}
\newcommand{\JCAP}{J. Cosmol. Astropart. Phys.}
\newcommand{\JHEP}{J. High. Energy. Phys.}
\newcommand{\bea}{\begin{eqnarray}}
\newcommand{\ena}{\end{eqnarray}}
\newcommand{\beqa}{\begin{eqnarray}}
\newcommand{\eeqa}{\end{eqnarray}}
\newcommand{\red}{\textcolor{black}}

\newlength\scratchlength
\newcommand\s[2]{
  \settoheight\scratchlength{\mathstrut}%
  \scratchlength=\number\numexpr\number#1-1\relax\scratchlength
  \lower.5\scratchlength\hbox{\scalebox{1}[#1]{$#2$}}%
}


\title{Parity-violating corrections to the orbital precession of binary system}

\author{Jin Qiao${}^{a,b}$}

\author{Qing-Guo Huang${}^{a,b,c}$}
\email{Corresponding author: huangqg@itp.ac.cn}

\author{Tao Zhu${}^{d, e}$}
\email{Corresponding author: zhut05@zjut.edu.cn}

\author{Wen Zhao${}^{f, g}$}

\affiliation{
${}^{a}$School of Fundamental Physics and Mathematical Sciences,
 Hangzhou Institute for Advanced Study, UCAS, Hangzhou 310024, China\\
${}^{b}$School of Physical Sciences, University of Chinese Academy of Sciences, No. 19A Yuquan Road, Beijing 100049, China\\
${}^{c}$CAS Key Laboratory of Theoretical Physics, Institute of Theoretical Physics,
 Chinese Academy of Sciences, Beijing 100190, China\\
${}^{d}$Institute for Theoretical Physics \& Cosmology, Zhejiang University of Technology, Hangzhou, 310023, China\\
${}^{e}$ United Center for Gravitational Wave Physics (UCGWP),  Zhejiang University of Technology, Hangzhou, 310023, China\\
${}^{f}$Department of Astronomy, University of Science and Technology of China, Hefei 230026, China \\
${}^{g}$School of Astronomy and Space Sciences, University of Science and Technology of China, Hefei, 230026, China}

\date{\today}

\begin{abstract}

In this work, we test for gravitational parity violation in the PSR J1141-6545 system by analyzing the orbital plane inclination precession induced by the misalignment between the white dwarf's spin axis and the system's total angular momentum. Using the parity-violating metric of gravity that incorporates terms from both the exterior and boundary of the field source, we calculated corrections to the relative acceleration and orbital inclination precession rates, which exhibit significant deviations from the GR prediction. The parity-violating contributions depend on the projection of the spin vector along the orbital angular momentum direction, contrasting with GR, where it depends on the projection within the orbital plane. The corrections are perpendicular to GR contribution, highlighting a fundamental distinction. The exterior field correction is linear in the theoretical parameter and coupled to eccentricity $e$, while the boundary term correction is quadratic. By comparing these corrections with GR and incorporating observational uncertainty, we derive the constraint on the theoretical parameter, yielding $ \dot{f}_{\rm PV}\lesssim 10^6~ \rm m$.

\end{abstract}

\maketitle

\section{Introduction}
\renewcommand{\theequation}{1.\arabic{equation}} \setcounter{equation}{0}

Since Einstein first proposed the theory of general relativity (GR) in 1915, research on gravitational interactions has advanced significantly, with GR demonstrating remarkable accuracy in numerous experiments and observations \cite{Will:2014kxa}. Symmetry is a fundamental feature of modern physical theories, providing a powerful framework for understanding gravitational interactions. Following the groundbreaking discovery of parity violation in weak interactions \cite{Lee:1956qn, Wu:1957my}, it became clear that nature does not always respect parity symmetry. However, GR is a fully symmetric theory and does not incorporate parity violation. To explore the possibility of parity violation in gravitational interactions, it is necessary to consider modified or alternative theories of gravity \cite{Alexander:2008wi}.

The most classical modified gravity theory incorporating parity violation is Chern-Simons (CS) modified gravity, which extends the metric beyond GR by introducing an additional term for the parity-violating Pontryagin density coupled to a scalar field \cite{Jackiw:2003pm, Yunes:2010yf, Yagi:2012vf,Alexander:2017jmt, Yagi:2017zhb,Alexander:2009tp}. However, CS gravity suffers from the Ostrogradsky instability due to its higher-order derivative field equations, limiting its applicability to a low-energy effective theory \cite{Woodard:2006nt}. To address this issue, CS gravity has been generalized to ghost-free parity-violating gravity by incorporating derivative terms of the scalar field \cite{Crisostomi:2017ugk}.
Simultaneously, a variety of modified and alternative gravitational theories have been proposed to explore the phenomenon of parity violation. These include Hořava-Lifshitz gravity \cite{Horava:2009uw,Wang:2012fi}, parity-violating spatial covariant gravity \cite{Yu:2024drx,Hu:2024hzo,Gao:2019liu}, Nieh-Yan modified teleparallel gravity \cite{Li:2021wij}, and symmetric teleparallel gravity within parity-violating frameworks \cite{Conroy:2019ibo,Li:2022vtn,Li:2021mdp}. Unlike conventional modified gravity theories, these alternative gravitational models are constructed within a non-Riemannian geometric framework, often utilizing non-metric or torsion descriptions of spacetime. These parity-violating gravities can produce observable phenomena that deviate from the predictions of GR.

Considering the presence of parity violation in gravity, the time-space component of the metric is corrected in the weak field limit using the parameterized post-Newtonian (PPN) approximation, introducing an additional PPN parameter \cite{Alexander:2007zg, Alexander:2007vt, Qiao:2021fwi, Rao:2021azn}. This component of the metric is closely related to the spin and angular momentum of the gravitational system, suggesting that the evolution of such dynamics can be used to test for parity violation in gravity. Naturally, parity violation would modify the frame-dragging effect predicted by GR, where spinning objects drag spacetime around them. In the solar system, the impact of parity violation on spin precession has been studied using data from the Gravity Probe B experiment, leading to constraints on the relevant parameters \cite{Alexander:2007zg, Smith:2007jm, Qiao:2021fwi}. Additionally, the influence of parity violation on orbital evolution has been investigated through observations of the LAGEOS and LARES satellites, providing further constraints on parity-violating parameters \cite{Smith:2007jm, Qiao:2023hlr}.

Compared to constraints derived from the solar system, which rely on the ratio of parity-violating corrections to the precession rates predicted by GR, gravitational waves (GWs) provide a more direct and robust avenue for probing parity-violating effects in gravity. The presence of parity violation modifies the dispersion relation and/or the friction term in the field equations, leading to velocity birefringence and/or amplitude birefringence during the propagation of GWs \cite{Zhao:2020}. For instance, Ref.~\cite {Nishizawa:2018srh, Zhao:2019szi} utilized velocity measurements from the GW170817 event to place bounds on the parity-violating parameter. Similarly, Ref.~\cite{Wang:2021gqm, Wang:2020cub, Wang:2025fhw, Zhu:2023rrx, Niu:2022yhr, Gong:2021jgg, Wu:2021ndf} investigated parity-violating effects in GW data by incorporating parity-violating waveforms, thereby deriving constraints on the relevant parameters in conjunction with observed GW events. Beyond GWs from isolated sources, parity violation in gravity can also be probed through B-mode polarization in the cosmic microwave background (CMB), which encodes information about primordial gravitational waves (PGWs)  \cite{Krauss:2010ty,Zhao:2014yna,Zhao:2014rya}. In the CMB, PGWs generate TT, EE, BB, and TE power spectra, whereas the TB and EB spectra vanish if the parity symmetry is preserved in gravity. The detection of nonzero TB and EB spectra at large-scale would provide strong evidence for parity violation in the gravitational sector \cite{Seto:2007tn,Lue:1998mq,Gluscevic:2010vv}. Furthermore, Refs.~\cite{Zhu:2022dfq, Li:2024fxy, Qiao:2019hkz, Zhu:2013fja, Wang:2012fi} explored the search for parity-violating signatures in the GW sector of the CMB polarization spectrum, offering additional insights into the nature of parity violation in the early universe.

In addition, the discovery of the first binary pulsar has provided a completely new testbed for gravitational tests \cite{Hulse:1974eb}. The discovery and continued observation of various binary pulsar systems have enabled the study of various aspects of gravitational interactions in strongly self-gravitating objects. Depending on the properties of the orbital evolution of the system and the characteristics of its companion star, these pulsar observations allow the testing of different aspects of gravity. Ref.~\cite{Yunes:2008ua} investigates the effect of parity violation on the orbital pericentre precession of binary pulsar systems, where the compact object is constant-density. Ref.~\cite{Ali-Haimoud:2011wpu} further investigates the effect of parity violation on the pericentre precession in binary system by considering more factors, thus giving constraints on the parity-violating parameter. 
The temporal evolution of the orbital inclination of this pulsar has recently been observed in the binary system PSR J1141-6545, which consists of a pulsar and a massive white dwarf companion \cite{VenkatramanKrishnan:2020pbi}.  The change in orbital inclination $\iota$ comes from orbital plane precession due to the spin of the white dwarf companion. The effect ultimately contributes to the temporal evolution of the observed projected semi-major axis, which was thus measured. This additional observed post-Keplerian parameter provides new opportunities to explore the parity violation in gravity. In this paper, we focus on the effect of parity violation on the temporal evolution of the orbital inclination. We use the observations to constrain the parity-violating parameter by investing the corrections to the orbital inclination precession.

This paper is organized as follows. In the next section, we briefly introduce parity-violating scalar-tensor gravity. In Section III, we describe the motion in the N-body system and compute the modified acceleration. In Section IV, we study the effect of the parity violation on the orbital evolution in binary systems and discuss constraints on the theoretical parameter. Conclusions and discussion are given in Section V.

Throughout this paper, the metric convention is chosen as $(-,+,+,+)$, and greek indices $(\mu,\nu,\cdot\cdot\cdot)$ run over $0,1,2,3$ and latin indices $(i, \; j,\;k)$ run over $1, 2, 3$.

\section{Ghost-free parity-violating gravity}
\renewcommand{\theequation}{2.\arabic{equation}} \setcounter{equation}{0}

The parity violation in gravitational interaction can arise from various beyond-GR theories, for example, ghost-free parity-violating gravity. In this subsection, we present a brief introduction to ghost-free parity-violating gravity. The action of the ghost-free parity-violating gravity has the following form \cite{Crisostomi:2017ugk}
\bqn\label{action}
\mathcal{S} = \f{1}{16 \pi G} \int d^4 x \sqrt{-g}\left[ R+\mathcal{L}_{\rm PV}+\mathcal{L}_{\phi} + \mathcal{L}_{\rm m}\right]\;,
\eqn
where \red{$G$ is Newton’s gravitational constant,} $R$ is the Ricci scalar, $\mathcal{L}_{\rm PV}$ is the Lagrangian which contains parity-violating terms coupled to a scalar field $\phi$, $\mathcal{L}_\phi$ is the Lagrangian for the scalar field and given by 
\bqn
\mathcal{L}_\phi =- \f{1}{2}g_{\mu\nu}\partial_{\mu}\phi\partial_{\nu}\phi-V(\phi),
\eqn
where $V(\phi)$ is the potential energy term.
 $\mathcal{L}_{\rm m}$ denotes the Lagrangian density of the matter field. The parity-violating Lagrangian $L_{\rm PV} $ has different expressions for different theories. The CS Lagrangian can be written in the form \cite{Alexander:2007zg}
\bqn
\mathcal{L}_{\rm CS} = \frac{1}{4} \vartheta(\phi) \;^*R R\;,
\eqn
where
\bqn
\;^*R R\;= \frac{1}{2} \varepsilon^{\mu\nu\rho\sigma} R_{\rho\sigma \alpha\beta} R^{\alpha \beta}_{\;\;\;\; \mu\nu}
\eqn
is the Pontryagin density defined with $\varepsilon_{\rho \sigma \alpha \beta}$ being the Levi-Civit\'{a} tensor defined in terms of the antisymmetric symbol $\epsilon^{\rho \sigma \alpha \beta}$ as $\varepsilon^{\rho \sigma \alpha \beta}=\epsilon^{\rho \sigma \alpha \beta}/\sqrt{-g}$ \red{and $\vartheta(\phi)$ is an arbitrary function of $\phi$}.  
The action \red{of CS theory} is generalized by including the first and second derivatives of the scalar field: $\phi_{\mu} = \partial_{\mu}\phi$ and $\phi_{\mu\nu} = \partial_{\mu}\phi_{\nu}$ \red{in Ref.~\cite{Crisostomi:2017ugk}}.

$\mathcal{L}_{\rm PV1}$ is the Lagrangian that contains the first derivative of the scalar field, which is given by \cite{Crisostomi:2017ugk}
\bqn\lb{LMPV1}
\mathcal{L}_{\rm PV1} &=& \sum_{\rm A=1}^4  a_{\rm A}(\phi, \phi^\mu \phi_\mu) L_{\rm A},\label{lv1}\\
L_1 &=& \varepsilon^{\mu\nu\alpha \beta} R_{\alpha \beta \rho \sigma} R_{\mu \nu}{}^{\rho}{}_{\lambda} \phi^\sigma \phi^\lambda,\nonumber\\
L_2 &=&  \varepsilon^{\mu\nu\alpha \beta} R_{\alpha \beta \rho \sigma} R_{\mu \lambda }^{\; \; \;\rho \sigma} \phi_\nu \phi^\lambda,\nonumber\\
L_3 &=& \varepsilon^{\mu\nu\alpha \beta} R_{\alpha \beta \rho \sigma} R^{\sigma}_{\;\; \nu} \phi^\rho \phi_\mu,\nonumber\\
L_4 &=&  \varepsilon^{\mu\nu\rho\sigma} R_{\rho\sigma \alpha\beta} R^{\alpha \beta}_{\;\;\;\; \mu\nu} \phi^\lambda \phi_\lambda,\nonumber
\eqn
with $\phi^\mu \equiv \nabla^\mu \phi$, and $a_{\rm A}$ are a priori arbitrary functions of $\phi$ and $\phi^\mu \phi_\mu$. In order to avoid the Ostrogradsky modes in the unitary gauge (where the scalar field depends on time only), it is required that $4a_1+2 a_2+a_3 +8 a_4=0$. With this condition, the Lagrangian in Eq. (\ref{lv1}) does not have any higher-order time derivative of the metric, but only higher-order space derivatives.

One can also consider the terms which contain second derivatives of the scalar field. Focusing on only these that are linear in Riemann tensor and linear/quadratically in the second derivative of $\phi$, the most general Lagrangian $\mathcal{L}_{\rm PV2}$ is given by \cite{Crisostomi:2017ugk}
\bqn\lb{LMPV2}
\mathcal{L}_{\rm PV2} &=& \sum_{\rm A=1}^7 b_{\rm A} (\phi,\phi^\lambda \phi_\lambda) M_{\rm A},\\
M_1 &=& \varepsilon^{\mu\nu \alpha \beta} R_{\alpha \beta \rho\sigma} \phi^\rho \phi_\mu \phi^\sigma_\nu,\nonumber\\
M_2 &=& \varepsilon^{\mu\nu \alpha \beta} R_{\alpha \beta \rho\sigma} \phi^\rho_\mu \phi^\sigma_\nu, \nonumber\\
M_3 &=& \varepsilon^{\mu\nu \alpha \beta} R_{\alpha \beta \rho\sigma} \phi^\sigma \phi^\rho_\mu \phi^\lambda_\nu \phi_\lambda, \nonumber\\
M_4 &=& \varepsilon^{\mu\nu \alpha \beta} R_{\alpha \beta \rho\sigma} \phi_\nu \phi_\mu^\rho \phi^\sigma_\lambda \phi^\lambda, \nonumber\\
M_5 &=& \varepsilon^{\mu\nu \alpha \beta} R_{\alpha \rho\sigma \lambda } \phi^\rho \phi_\beta \phi^\sigma_\mu \phi^\lambda_\nu, \nonumber\\
M_6 &=& \varepsilon^{\mu\nu \alpha \beta} R_{\beta \gamma} \phi_\alpha \phi^\gamma_\mu \phi^\lambda_\nu \phi_\lambda, \nonumber\\
M_7 &=& (\nabla^2 \phi) M_1,\nonumber
\eqn
with $\phi^{\sigma}_\nu \equiv \nabla^\sigma \nabla_\nu \phi$. Similarly, in order to avoid the Ostrogradsky modes in the unitary gauge, the following conditions should be imposed: $b_7=0$, $b_6=2(b_4+b_5)$ and $b_2=-A_*^2(b_3-b_4)/2$, where $A_*\equiv \dot{\phi}(t)/N$ and $N$ is the lapse function. In this paper, we consider a general scalar-tensor theory with parity violation, which contains all the terms mentioned above. So, the parity-violating term in Eq.(\ref{action}) is given by
\bqn
\mathcal{L}_{\rm PV} = \mathcal{L}_{\rm CS} + \mathcal{L}_{\rm PV1} + \mathcal{L}_{\rm PV2}.
\eqn
Therefore, the CS modified gravity in \cite{Alexander:2007zg}, and the ghost-free parity-violating gravities discussed in  \cite{Crisostomi:2017ugk} are all the specific cases of this Lagrangian. The \red{coefficients  $a_{\rm A}$ and $b_{\rm A}$ depend on the scalar field $\phi$ and its evolution.}
The field equations for this modified gravity theory have been explicitly derived in Ref. \cite{Qiao:2021fwi}.

\section{Motion of compact bodies}
\renewcommand{\theequation}{3.\arabic{equation}} \setcounter{equation}{0}
\lb{sec3}

This section begins by presenting the equations of motion of the system when the fluid splits into multiple separated components, i.e., the post-Newtonian N-body problem. The equation of motion of the system is derived by employing a covariant expression for the conservation of the energy-momentum tensor and calculating the energy-momentum tensor with the required accuracy, which has been widely used in the investigation of binary or N-body systems in astronomy.
According to Ref.~\cite{will2018th} we first present the method briefly, assuming that a form of  the metric to post-
Newtonian order as
\bqn\lb{meq1}
g_{00} &=& -1 + \f{2}{c^2}U+ \f{2}{c^4}U(\Psi-U^2) + O(c^{-6}),\nb\\
g_{0i} &=& -\f{4}{c^2}U_i  + O(c^{-5}),\nb\\
g_{ij} &=& \left(1 + \f{2}{c^2}U \right)\delta_{ij} + O(c^{-6}),
\eqn
where $U$, $U_i$, and $\Psi$ are gravitational potentials to be determined.

The energy-momentum tensor of a perfect fluid is 
\bqn
T^{\alpha\beta}= \left(\rho + \f{\rho\Pi}{c^2}+\f{p}{c^2} \right)u^{\alpha}u^{\beta}+pg^{\alpha\beta},
\eqn
where $\rho$ is the proper mass density, $\Pi$ is the proper internal energy, $p$ is the pressure, and $u^{\alpha}$ is the velocity field.
The equation of energy-momentum conservation is 
\bqn
0= \nabla_{\beta}T^{\alpha\beta}=\partial_{\beta}T^{\alpha\beta}+\Gamma^{\alpha}_{\mu\beta}T^{\mu\beta}\red{+\Gamma^{\beta}_{\mu\beta}T^{\alpha\mu}},
\eqn 
applying the relation 
\bqn
\Gamma^{\beta}_{\mu\beta}=\f{1}{2}g^{\beta\nu}\partial_{\mu}g_{\beta\nu}=(-g)^{-1/2}\red{\partial_{\mu}}(-g)^{1/2}
\eqn
simplifies to
\bqn\lb{emq}
0=\partial_{\beta}(\sqrt{-g}T^{\alpha\beta})+\Gamma^{\alpha}_{\beta\gamma}(\sqrt{-g}T^{\beta\gamma}).
\eqn
The squre root of the metric determinant is given by $\sqrt{-g}= 1+2 c^{-2}U + \mathcal{O}(c^{-4})$. The components of the energy-momentum tensor are given by
\bqn
c^{-2} T^{00} &=& \rho \left[ 1+ \f{1}{c^2}\left( \f{1}{2}v^2 - U + \Pi \right) \right] + \mathcal{O}(c^{-2}),\nb\\
c^{-1} T^{0j} &=& \rho v^j \left[ 1+ \f{1}{c^2}\left( \f{1}{2}v^2 - U + \Pi + \f{p}{\rho}\right) \right]\nb\\
&&+ \mathcal{O}(c^{-4}),\nb\\
T^{ij} &=& \rho v^i v^j \left[ 1+ \f{1}{c^2}\left( \f{1}{2}v^2 - U + \Pi + \f{p}{\rho}\right) \right] \nb\\
&&+ p(1-\f{2}{c^2}U) \delta^{ij} + \mathcal{O}(c^{-4}).
\eqn
\subsection{Energy conservation}

\red{The zeroth component of Eq.~(\ref{emq})} gives rise to a statement of energy conservation. The expanded equation is 
\bqn
0& =& \f{1}{c} \partial_t(\sqrt{-g}T^{00}) + \partial_j(\sqrt{-g}T^{0j}) + \Gamma^{0}_{00}(\sqrt{-g}T^{00}) \nb\\
&&+ 2\Gamma^0_{0j}(\sqrt{-g}T^{0j}) + \Gamma^0_{jk}(\sqrt{-g}T^{jk}). 
\eqn 
After inserting the components of $T^{\alpha\beta}$ and $\Gamma^{\alpha}_{\mu\nu}$, at order $c$ \red{in} the continuity equation 
\bqn\lb{eqlx}
0 &=& \partial_t\rho + \partial_j(\rho v^j). 
\eqn
At order $c^{-1}$ have 
\bqn
0&=& \rho \partial_t\left( \f{1}{2} v^2  + \Pi \right) + \rho v^j\partial_j\left( \f{1}{2} v^2  + \Pi \right)\nb\\
&&+ \partial_j (pv^j) - \rho v^j \partial_j U .
\eqn
\red{This} equation expresses the local conservation of energy within the fluid.

\subsection{Momentum conservation}

The spatial components of Eq.~(\ref{emq}) provide a statement of momentum conservation and are expressed as
\bqn
0 &=& \f{1}{c}\partial_t(\sqrt{-g}T^{0j}) + \p_k(\sqrt{-g}T^{jk})+ \Gamma^j_{00}(\sqrt{-g}T^{00})\nb\\
&&+ 2\Gamma^j_{0k}(\sqrt{-g}T^{0k}) + \Gamma^j_{kn}(\sqrt{-g}T^{kn}).
\eqn
After some algebra and simplification, the equation becomes
\bqn\lb{eqpave}
\rho\f{dv^j}{dt} &=& -\partial_jp + \rho\partial_jU \nb\\
&&+ \f{1}{c^2}\left[- v^j\p_tp + \frac{1}{c^2}\left( \f{1}{2}v^2 +U + \Pi +\f{p}{\rho}\right)\partial_jU\right]\nb\\
&&+ \f{1}{c^2}\rho\big[(v^2-4U)\partial_jU - v^j(3\p U+4v^k\p_kU)\nb\\
&&+ 4\p_tU_j + 4v^k(\p_kU_j - \p_jU_k) + \p_j\Psi \big]+\mathcal{O}(c^{-4}).\nb\\
\eqn
This equation is the post-Newtonian version of Euler's equation.
Generally this equation, together with the continuity equation (\ref{eqlx}) and the equation of state, provides a complete description of the dynamical behaviour of a slowly moving fluid in a weak gravitational field.

The equations of motion for the center-of-mass positions $\bl{r}_A(t)$ can be written as
\bqn
m_A\bl{a}_A = \int_A\rho\f{d\bl{v}}{dt} d^3x.
\eqn
Now, we only need to determine the expression for the gravitational potential and substitute Eq. (\ref{eqpave}) into the above equation, where the expression for the acceleration can be explicitly derived through intricate integration and meticulous mathematical manipulation.
The definitions of Newtonian and post-Newtonian potentials included in the metric we list \red{in the Appendix (\ref{App}-\ref{App1})}. In the following analysis, we focus on the influence of the rotating body's spin on the system's motion. Consequently, only the expressions for the spin-dependent component of the potential are provided. The spin angular momentum tensor and vector of a rotating body are defined as follows, respectively:
\bqn
S^{ij}_A &\equiv& \int_A\rho(x^iv^j-x^jv^i)d^3x;\nb\\
\bl{S}&\equiv &\int_A\rho ~\bl{x}\times \bl{v}d^3x;
\eqn
which are related to each other as
\bqn
S^i = \f{1}{2} \epsilon^{ijk}S^{jk}_A,~~~S^{ij}_A = \epsilon^{ijk}S^k_A.
\eqn
The expressions for the spin-dependent component of the potential are
\bqn
U^i &=& -\f{1}{2}\sum_A\f{GS^{ij}_An^j_A}{x^2},\nb\\
\Psi &=& -\f{3}{2}\sum_A\f{Gv^i_AS^{ij}_An^j_A}{x^2},
\eqn
where $\bl{x}$ is the position of a fluid element relative to the body's center-of-mass, and $n^i\equiv {\bl{x}}/{x}$ defines the unit vector in the direction of $\bl{x}$.
In \red{GR, the contribution of the spin-orbit coupling to }the acceleration of a rotating body, $\red{{a}^j_A[\rm SO] }$, is finally calculated as \cite{will2018th}
\bqn\lb{so1}
\red{{a}^j_A[\rm SO] }&=& \f{3}{2c^2} \sum_{A\neq B}\f{Gm_B}{r^3_{AB}}\Big\{ n^{\langle jk \rangle}_{AB} \nb\\
&& \left[ v^p_A\left( 3 \hat{S}^{kp}_A + 4\hat{S}^{kp}_B \right) - v^p_B\left( 3 \hat{S}^{kp}_B + 4\hat{S}^{kp}_A \right)\right]\nb\\
&&+n^{\langle kp \rangle}_{AB} \left( v_A - v_B \right)^p \left(  3 \hat{S}^{jk}_A + 4\hat{S}^{jk}_B  \right)\Big\},
\eqn
which $\hat{S}^{jk}_i \equiv  {S}^{kp}_i/m_i$, ${S}^{jk}_i \equiv  \epsilon^{jkp}{S}^{p}_i$. $n^{\langle jk \rangle}_{AB}$  denotes a symmetric traceless relational equation as
\bqn
n^{\langle jk \rangle}_{AB} &\equiv& n^j_{AB}n^k_{AB} - \frac{1}{3}\delta^{jk}.
\eqn
\red{Eq.~(\ref{so1}) presents the contribution of the spin-orbit coupling to the acceleration in GR, which is linear in each spin tensor.
Furthermore, rotating celestial bodies in the system exert mutual influences, giving rise to spin-spin acceleration. This contribution is not considered in the present work, as spin-spin acceleration is a secondary effect compared to spin-orbit acceleration. In the observational effects of the binary system PSR J1141-6545 that we will apply below, only the contribution from spin-orbit interaction has been detected so far. Therefore, we neglect the contribution of spin-spin interaction to the acceleration here.} As established by the action in Eq.~(\ref{action}), the parity-violating terms under investigation are introduced as additional contributions to the GR framework. Consequently, it is natural to infer that these parity-violating terms will generate the new correction terms in the acceleration Eq. (\ref{so1}). This correction will be discussed in detail below.

\subsection{Spin-orbit acceleration from parity-violating terms}

The acceleration derived from the post-Newtonian N-body equations of motion in GR has been previously established. In this study, we investigate how these N-body accelerations are modified when parity-violating terms are introduced to correct the metric. Specifically, within the framework of parity-violating gravity, we focus on a single source characterized solely by rotational dynamics. Using the PPN formalism, the metric correction term for GR in this context is given by \cite{Alexander:2007zg}
\bqn\lb{cs1}
\delta g_{0i} = \dot{f}_{\rm PV} \left[ -\f{S^i}{r^3} + 3\f{S^jn^j}{r^3}n^i \right],
\eqn
where\red{ $f_{\rm PV}  = \vartheta(\phi) + (- 2a_2 + a_3 -8 a_4)\dot{\phi}^2$.}
This correction term reflects the additional contributions arising from parity-violating effects, which are not accounted for in the standard GR framework. Our analysis seeks to quantify the impact of these corrections on the N-body dynamics and explore their implications for gravitational interactions in systems where parity violation may play a significant role.
The solution to the field equations is derived as a lower-order approximation by treating the source as a point particle. However, this approximation only describes the external gravitational field of the source and fails to ensure continuity of the field as it transitions through the surface of the sphere \cite{Alexander:2007zg,Smith:2007jm}. 

With the correction to the metric given by Eq. (\ref{cs1}), we can follow the same procedure as before to derive the corresponding correction to the N-body acceleration. This correction takes the form
\bqn\lb{cs2}
\delta a^i_{A}
&=&-\f{1}{c^2}  \ddot{f}_{\rm PV} \sum_{A\neq B}\left[ -\f{S_B^i}{r^3} + 3\f{S_B^jn_{12}^j}{r^3}n^i_{12} \right]\nb\\
&&+ \f{1}{c^2}\dot{f}_{\rm PV} \sum_{A\neq B} \f{1}{r^4}  \Big[ 3 v_2^kS_B^i n^k_{12} \nb\\
&&+ 3v_2^kS_B^k n^i_{12} +  3 v_2^iS_B^j n^j_{12}  - 15 v_2^kS_B^jn^j_{12}n^i_{12}n^k_{12} \Big]. \nb\\
\eqn 
From the above expression, it is evident that the correction to the N-body acceleration induced by the parity-violating term originates solely from the spin effect of the external source $B$. In contrast, within the GR, both the resultant source $A$ and the external source 
$B$ simultaneously contribute to the acceleration. Our calculations further reveal that for body $A$, the spin-orbit effects are mutually canceling and thus do not contribute to the acceleration. This highlights a key distinction between the parity-violating correction and the standard GR contributions, emphasizing the unique role of spin effects in modifying gravitational dynamics. 
\red{It is worth noting that, similar to Eq. (\ref{so1}), Eq. (\ref{cs2}) is also only valid at the linear order in spin.}

\section{Binary systems}
\renewcommand{\theequation}{4.\arabic{equation}} \setcounter{equation}{0}
\lb{sec4}

We now consider a binary system in which two objects have masses $m_1$ and $m_2$, velocities $\bl v_1$ and $\bl{v}_2$, and positions $\bl{r}_1$ and $\bl{r}_2$.
We define the separation vector and the relative velocity of the binary system as respectively
\bqn
\bl{r} \equiv \bl{r}_1 - \bl{r}_2, ~~~\bl{v} \equiv \bl{v}_1 - \bl{v}_2,
\eqn
and we set 
\bqn
r \equiv |\bl{r}|,~~~ \bl{n} \equiv \f{\bl{r}}{r}, ~~~{v} \equiv |\bl{v}|.
\eqn

\subsection{Relative acceleration in binary systems}

 According to the spin-orbit acceleration of the Eq. (\ref{so1}), the expression for the relative spin-orbit acceleration 
$\bl{a}\equiv \bl{a}_1 - \bl{a}_2$ of a binary star system reads
\bqn\lb{GR1}
a^j &=& \f{3G}{2c^2r^3} \Big\{ n^{\langle jk \rangle}v^p\left[3\sigma^{kp} + 4S^{kp}\right] \nb\\
 &&  n^{\langle kp \rangle} v^p\left[3\sigma^{jk} + 4S^{jk}\right] \Big\},
\eqn 
 which translates into a vector of the form
\bqn
 \bl{a}&=& \f{3G}{2c^2r^3} \Big\{ \bl{n}(\bl{n}\times\bl{v})\cdot(4\bl{S+3\bl{\sigma}})\nb\\
 &&+ (\bl{n}\cdot\bl{v})\bl{n}\times(4\bl{S+3\bl{\sigma}}) 
  -\f{2}{3} \bl{v}\times (4\bl{S+3\bl{\sigma}})\Big\},\nb\\
\eqn
in which 
\bqn
\bl{\sigma} \equiv \f{m_2}{m_1}\bl{S}_1 + \f{m_1}{m_2}\bl{S}_2,~~~\bl{S}\equiv \bl{S}_1 + \bl{S}_2.
\eqn
The following relationship is used in the expression of relative acceleration
\bqn
\bl{v}_1 = \f{m_2}{m}\bl{v},~~~\bl{v}_2 = -\f{m_1}{m}\bl{v},~~~m\equiv m_1+m_2.
\eqn
In the binary system according to the modified acceleration Eq. (\ref{cs2}), we give the relative modified acceleration $\delta\bl{a}\equiv \delta\bl{a}_1 - \delta\bl{a}_2$ as
\bqn\lb{cs3}
\delta a^i&=& \f{1}{c^2}  \ddot{f}_{\rm PV} \f{1}{r^3} \left[  S_2^i -  S_1^i  + 3 S_1^jn^j n^i -3 S_2^jn^j n^i\right]\nb\\
&&+ \f{1}{c^2}\dot{f}_{\rm PV}  \f{1}{r^4}  \Big[- 3 \f{m_1}{m}v^kS_2^i n^k - 3\f{m_1}{m}v^kS_2^k n^i \nb\\&&
- 3 \f{M_1}{m}v^iS_2^j n^j + 15 \f{m_1}{m}v^kS_2^jn^jn^in^k \nb\\
&&+ 3 \f{m_2}{m}v^kS_1^i n^k  + 3\f{m_2}{m}v^kS_1^k n^i  \nb\\
&&+  3 \f{m_2}{m}v^iS_1^j n^j  - 15 \f{m_2}{m}v^kS_1^jn^j n^i n^k \Big].
\eqn
\red{In the above expressions, we have derived the correction to the relative acceleration induced by parity violation at the linear order of spin. Analysis shows that in this corrected relative acceleration, the second-order time derivative of the theoretical parameter $\ddot{f}_{\rm PV}$ is directly coupled to the spin, while its first-order time derivative $\dot{f}_{\rm PV}$ is coupled to both the velocity and the spin. Using this corrected relative acceleration, we can directly calculate the orbital elements and further obtain the contribution of the parity-violating term to the observable quantities. A detailed discussion on this will be presented in the following subsections.
}

\subsection{Evolution of orbital inclination}

The binary system is subject to spin-orbit effects, which will lead to different changes in orbital evolution. Calculations in GR show that the spin-orbit interaction will produce Lense-Thirring precession, which cumulates the orbital elements in the secular evolution.
The temporal evolution of a pulsar's orbital inclination has been observed and is inferred to be due to a combination of Newtonian quadrupole moments and Lense-Thirring orbital precession due to the fast rotation of the white dwarf. Here we are interested in how the presence of the parity-violating terms affect the evolution of the orbital inclination. In the following, we will specifically calculate the results of the parity-violating correction.

The positions and velocities of the orbits are defined in respect of the orbital elements as follows
\bqn
\bl{r} &\equiv& \f{p}{1+e\cos \theta}\bl{n},\nb\\
\bl{v} &\equiv& \dot{r} \bl{n} + \f{h}{r} \bl{\lambda},\nb\\
h &\equiv& \sqrt{Gmp},
\eqn
 where $p$ is the semi-latus rectum of the elliptical motion, $e$ is the orbitial eccentricity, and $f$ is the true anomaly. The unit vectors $\bl{n}$ and $\bl{\lambda}$ denote the radial and transverse direction respectively. With $\bl{n}$ and $\bl{\lambda}$ we can introduce the
 unit normal vector $\hat{\bl{h}} = \bl{n} \times \bl{\lambda}$.
 
 The motion of binary system can be treated as perturbations to the Kepler problem in Newtonian mechanics. The equations of motion are equivalent to the equations of the osculating orbital elements.
 Here we are only concerned with the change in orbital inclination, the equation for orbital inclination can be given \cite{Poisson2024}
\bqn\lb{w}
\f{d\iota}{d f} &\simeq& \f{p^2}{Gm}\f{\cos(\omega+f)}{(1+ e\cos f)^3}\mathcal{W},
\eqn
{where $\mathcal{W}$ denotes the projection of the acceleration in the $\hat{\bl{h}}$ direction.}
We find the projections of Eq. (\ref{GR1}) and Eq. (\ref{cs3}) in the $\hat{\bl{h}}$-direction respectively as
\bqn
\mathcal{W} &=& \bl{a}\cdot \hat{\bl{h}}\nb\\
&=& \frac{1}{c^2}  \frac{G(1+e\cos f)^3}{p^3} \Bigg\{ 
  \f{1}{2} \sqrt{\f{Gm}{p}}e\sin f \nb\\
  &&(-\sin f\bl{e}_x + \cos f\bl{e}_y)   \cdot \left( 4  {\bm S} +3 {\bm{\sigma}} \right)    \nb\\
&& +\f{\sqrt{Gmp}(1+e\cos f)}{p}\nb\\
&&(\cos f\bl{e}_x + \sin f\bl{e}_y) \cdot \left( 4  {\bm S} +3 {\bm{\sigma}} \right)  \Bigg\},\nb\\
\delta\mathcal{W} &=& \delta\bl{a}\cdot \hat{\bl{h}}\nb\\
&=&\f{G}{c^2}  \ddot{f}_{\rm PV} \f{1}{r^3} \left[  \bl{S}_2 -  \bl{S}_1   \right]\cdot \hat{\bl{h}}\nb\\
&&+ \f{G}{c^2}\dot{f}_{\rm PV}  \f{1}{r^4}  \Big[- 3 \f{m_1}{m}(\bl{v}\cdot\bl{n})\bl{S}_2   \nb\\
&&+ 3 \f{m_2}{m}(\bl{v}\cdot\bl{n})\bl{S}_1  \Big]\cdot \hat{\bl{h}}.
\eqn
We insert the above equation into the orbit inclination element Eq.(\ref{w}) and average with integration over the interval $[0,2\pi]$ of one orbital period. Finally we get
\bqn
\f{d\iota}{d t} &=& \f{1}{T} \int^{2\pi}_{0}df\f{d\iota}{d f} \nb\\
&=& \frac{1}{2c^2} \f{G}{a^{3}(1-e^2)^{3/2}}  \Bigg[ \left(4 +3 \frac{m_2 }{m_1}\right )  \bl{I} \cdot \bl{ S}_1  \nb\\
&&+ \left(4+3\frac{m_1 }{m_2}\right)  \bl{I} \cdot  \bl{S}_2  \Bigg] , \lb{dm1} \\
\delta \f{d\iota}{d t} &=& \f{1}{T} \int^{2\pi}_{0}df \delta \f{d\iota}{d f} \nb\\
&=&   -   \f{G}{2c^2}  \f{\dot{f}_{\rm PV}e}{a^4(1-e^2)^{5/2} } \sin\omega  \nb\\
&&\Big[- 3 \f{m_1}{m} \bl{S}_2 \cdot \hat{\bl{h}}   + 3 \f{m_2}{m}\bl{S}_1 \cdot \hat{\bl{h}} \Big],\lb{dm2}
\eqn
where  $\bl{I} = \cos \omega   \bl{e}_x -\sin \omega   \bl{e}_y $ is a unit vector. These two expressions are the GR and parity-violating terms that produce orbital inclination variations in a period, of which the first formula was first given by Damour and Schäfer \cite{Damour:1988mr}. From Eq. (\ref{dm1}) and Eq. (\ref{dm2}) we can see that the first expression shows that the variation of the orbital inclination in GR is related to the contribution of the spin vector projected in the orbital plane. On the other hand, the second expression is the parity-violating contribution, which is mainly related to the projection of the spin in the direction of the total orbital angular momentum.
It is important to note that our procedure above does not give a decomposition of the spin vectors. If the spin vectors are assumed to be parallel to the direction of the total orbital angular momentum, then the first expression above are zero. In binary systems, if the two spin vectors are not parallel to the direction of the total orbital angular momentum of the binary system, the total angular momentum must remain conserved (up to the 2PN approximation), which leads to a corresponding change in the direction of the orbital angular momentum when the spin direction changes \cite{Damour:1988mr,VenkatramanKrishnan:2020pbi,Freire:2024adf}. This effect can be measured by experimental observations, which can then be applied to test gravity.

\subsection{Contribution of orbital inclination in observations}
\red{PSRJ1141-6545 is a radio pulsar with a spin period of 394 ms in a 4.74 hr eccentric orbit, and it has a massive WD companion.}
PSR J1141-6545 has been continuously observed by researchers since 2000. The observing team has given measurements of some of the PK parameters through sustained observations (among them periastron advancement, relativistic time dilation, gravitational wave damping, and the Shapiro delay) all of which show that the observational data for these parameters are in excellent agreement with the GR. 
Recently, the observing team has given additional measurement of the PK parameter, that is the temporal evolution of the projected-semi major axis $x_{\rm obs}$.

The expression for projected-semi major axis $x_{\rm obs}$ in a binary system with the necessary accuracy can be written as
\bqn\lb{obs1}
x_{\rm obs} = \f{a}{c}\sin \iota + A,
\eqn
where $A$ describes how the aberration of the pulsar's signal affects the measurement of $x$. 
Current research suggests that generally in binary pulsar systems, the observed variation in the projected semi major axis of the pulsar orbit, $\dot{x}_{\rm obs}$, may be due to a variety of physical and geometrical factors, which can be decomposed as follows
\bqn
\dot{x}_{\rm obs} = \dot{x}_{\rm PM}+\dot{x}_{\rm \dot{D}}+\dot{x}_{\rm GW}+\dot{x}_{\rm \dot{m}}+\dot{x}_{\rm 3rd}+\dot{x}_{\rm \epsilon_{\dot{A}}}+\dot{x}_{\rm SO},\nb\\
\eqn
where $\dot{x}_{\rm PM}$ is the proper motion of the system, $\dot{x}_{\rm \dot{D}}$ is the changing radial Doppler shift, $\dot{x}_{\rm GW}$ is gravitational wave emission, $\dot{x}_{\rm \dot{m}}$ is mass-loss in the system, $\dot{x}_{\rm 3rd}$ is the presence of a hypothetical third body in the system, $\dot{x}_{\rm \epsilon_{\dot{A}}}$ is a secular change in the aberration of the pulsar beam due to geodetic precession, and $\dot{x}_{\rm SO}$ is spin-orbit coupling, \red{ with contributions from mass quadrupole moments $\dot{x}_{\rm QPM}$ and Lense-Thirring orbital precession $\dot{x}_{\rm LT}$}. 
The present observational investigation in Ref. \cite{VenkatramanKrishnan:2020pbi} suggests that only two of these effects, $\dot{x}_{\rm \epsilon_{\dot{A}}}$ and $\dot{x}_{\rm SO}$, have a dominant contribution on the same order of magnitude,
meanwhile $\dot{x}_{\rm \epsilon_{\dot{A}}}$ contributes less than $21\%$ to the $x_{\rm obs}$ at $99\%$ confidence level.

In this section, we are concerned with the parity-violating corrections to the prediction in GR, which depend exclusively on the spin-orbit coupling $\dot{x}_{\rm SO}$. We mainly analyze the constraints on the parity-violating parameter provided by the observed effect of $\dot{x}_{\rm SO}$. In the binary system of PSR J1141-6545, we assume a spin of $\bl{S}_1$ for the pulsar and a spin of $\bl{S}_2$ for the massive white dwarf companion.
 The temporal evolution of the projected semi-major is due to Newtonian quadrupole moments $\dot{x}_{\rm QPM}$ and Lense-Thirring orbital precession $\dot{x}_{\rm LT}$ resulting from rapid rotation of the white dwarf. $\dot{x}_{\rm QPM}$ and $\dot{x}_{\rm LT}$ are simultaneously modulated by the spin misalignment angle $\delta_c$ and the precession phase $\Phi_c$ of the white dwarf during selected periods. 
 The Lense-Thirring precession is always present throughout the evolution of the system, regardless of the choice of the two parameters $\{\delta_c, \Phi_c\}$ within a reasonable range of values. Ref. \cite{VenkatramanKrishnan:2020pbi} uses simulations to give the relationship between $\dot{x}_{\rm QPM}$ and $\dot{x}_{\rm LT}$  as a function of the white dwarf's spin period $P_{\rm GW}$. In order to consider the effect of the parity-violating correction in the Lense-Thirring precession,  the extreme case may be chosen here: \red{ the  Lense-Thirring precession dominates when $P_{\rm WD}>270~ \rm s$; then the effect of $\dot{x}_{\rm QPM}$ can be ignored. Ref. \cite{VenkatramanKrishnan:2020pbi} reveals that when the spin period $P_{\rm WD} < 200~ \rm s$, the contribution from the mass quadrupole moment exhibits an opposite sign to that of the LT precession, leading to a partial cancellation of the LT effect. This cancellation mechanism motivates our adoption of the condition $P_{\rm WD} > 270~ \rm s$, as established in Ref. \cite{VenkatramanKrishnan:2020pbi}, where the LT precession dominates the dynamical evolution.}

 We began by considering the LT precession of the orbital plane, whose main contribution comes from the spin of the white dwarf companion. 
 \red{Since the variation rate induced by the pulsar's spin moment is smaller than the measurement error for nearly all physically plausible parameter values, this contribution can be safely neglected in Eq.(\ref{dm1}-\ref{dm2}). 
 In the PSR J1141-6545 system, the maximum values of the pulsar's mass quadrupole moment and angular momentum are $Q_2\leq 3.1 \times 10^{37}~\rm kg~ m^2$ and $S_2\leq 4 \times 10^{40}~\rm kg~ m^2s^{-1}$, respectively \cite{Iorio:2020xos}. By comparing the maximum absolute values of their contributions to the pulsar's observable quantity $\dot{x}^{P}_{\rm SO}$, we find that the resulting observable quantity $\dot{x}^{P}_{\rm SO}\leq 2.8\times 10^{-14} ~\rm ss^{-1} $ \cite{Iorio:2020xos} of the pulsar is actually slightly smaller than the uncertainty of the observable quantity $0.3\times 10^{-13}~\rm ss^{-1}$.
}
 Neglecting the spin contribution of the pulsar, from Eq. (\ref{dm1}) and Eq. (\ref{obs1}) we have \cite{Damour:1988mr,VenkatramanKrishnan:2020pbi,Freire:2024adf}
\bqn\lb{x12}
\dot{x}_{\rm LT} &=& -x \f{GS_2}{c^2a^{3}(1-e^2)^{3/2}} \nb\\
&&\times \left(2+\frac{3m_1 }{2m_2}\right) \cot\iota\sin \delta_c\sin \Phi_c  ,\nb\\
\delta\dot{x}_{\rm LT} &=& -x   \f{\dot{f}_{\rm PV}eGS_2}{c^2 a^4(1-e^2)^{5/2} } \sin\omega   \f{3m_1}{2m} \cot\iota \cos \delta_c.\nb\\
\eqn
The first expression above is the contribution to the temporal evolution of $x$ from the Lense-Thirring precession given in GR, and the second expression is the correction from the parity-violating terms we are considering. 
Note that the main contribution of the second expression scales with $e$, whereas the eccentricity is small but does not vanish in the system discussed here. The reason for the scaling of $e$ occurs is that the terms of $e$ do not vanish upon integration. In contrast to this it has been shown in Ref. \cite{Yunes:2008ua} that CS term scales with $1/e$ in the contribution of $\dot{w}$. \red{ Calculations indicate that both Eq. (\ref{x12}) and $\dot{w}$ contain the eccentricity $e$. In a binary system, the eccentricity satisfies $e<1$; consequently, the eccentricity in Eq. (\ref{x12}) suppresses the contribution of the parity-violating terms, whereas $1/e$ in $\dot{w}$ acts to enhance the contribution of the parity-violating terms \cite{Yunes:2008ua}.}

Timing observations of PSR J1141-6545 have been conducted using the 64-meter Parkes radio telescope and the UTMOST telescope. Based on these observations, we have presented a set of measured and derived parameters for the system, which are summarized in Table \ref{tab:no_border} \cite{VenkatramanKrishnan:2020pbi}.
\begin{table}[htbp]
    \centering
    \caption{Model parameters for PSR J1141-6545 \cite{VenkatramanKrishnan:2020pbi}.}
    \label{tab:no_border}
    \begin{tabular}{lcc} 
        \hline\hline 
       Measured quantities \\ \hline 
        Projected semi-major axis,$x (\rm s)$   & 
        $1.858915 \pm3 \times 10^{-6}$   \\
       Longitude of periastron, $\omega_0$ (deg)    &
       $ 80.6911 \pm 6 \times 10^{-4}$   \\
       Orbital eccentricity, $e$  & $0.171876 \pm 1 \times 10^{-6}$\\
       First derivative of $x$, $\dot{x} (\rm s s^{-1})$  & $(1.7\pm 0.3) \times 10^{-13}$ \\
       Companion mass, $m_c (M_{\odot} )$ & $1.02\pm 0.01$\\
       Total Mass, $m_{\rm TOT} (M_{\odot})$ & $2.28967 \pm 6 \times 10^{-5}$\\ \hline
       Derived quantities\\ \hline
       Pulsar mass, $m_p (M_{\odot} )$ &$1.27\pm 0.01$\\ 
       Orbital inclination, $\iota$ (deg) &  $71 \pm 2$ or $109 \pm 2$\\
        \hline \hline
    \end{tabular}
\end{table}
The current observations give a measurement of $\dot{x}_{\rm obs}$ as $\dot{x}_{\rm obs} = (1.7 \pm 0.3)\times 10^{-13} \rm ss^{-1}$ \cite{VenkatramanKrishnan:2020pbi}, where the contribution from spin-orbit interactions $\dot{x}_{\rm SO}$ dominates the contribution by more than $79\%$, which in the extreme case we are considering can well be regarded as a contribution from the Lense-Thirring precession.
From the observation of the projected semi-major axis, we deduce that 
$a = 5.3\times 10^5 \rm km$, and this result is consistent to that level of accuracy regardless of which derived inclination value is used.

Therefore the parity-violating contribution we consider must be smaller than the error bounds of this measurement. With Eq. (\ref{x12}) we give the ratio of the parity-violating contribution to the Lense-Thirring precession in GR as
\bqn\lb{eq4.16}
\f{\delta\dot{x}_{\rm LT}}{\dot{x}_{\rm LT} } = \f{\dot{f}_{\rm PV}e}{a(1-e^2)}\f{3m_1m_2}{m(4m_2+3m_1)}\sin\omega\cot\delta_c\cos\Phi_c.\nb\\
\eqn
Combined with the observations we can obtain from this correction ratio a constraint on the relationship of parity-violating parameter which is expressed as
\bqn\lb{geq}
 \dot{f}_{\rm PV} &<& \f{ a(1-e^2)}{e}\f{m(4m_2+3m_1)}{3m_1m_2}\red{\f{\tan\delta_c}{\sin\omega\cos\Phi_c}}\nb\\
 &&\times\f{0.3}{1.7\times79\%},
\eqn
\red{where in the order-of-magnitude estimation, the trigonometric functions related to $\delta_c$ and $\Phi_c$ are approximated as $\mathcal{O }(1)$ (i.e., their values can be roughly regarded as constants of order 1). The precession of the spin vector $\bl{S}_2$ induces the corresponding changes in angles $\delta_c$ and $\Phi_c$; this precession, which arises from the contribution of orbital precession, is a small quantity that accumulates over time. Eq. (\ref{eq4.16}) represents the ratio of the corrected parity-violating LT precession to GR precession within one orbital period $P_{\rm b} = 4.74~ \rm hr$ for SR J1141-6545. When performing the final order-of-magnitude estimation based on Eq. (\ref{geq}), the changes in the two parameters $\delta_c$ and $\Phi_c$ from the initial to the final moment within one orbital period can be safely neglected, without significantly affecting the rationality of the estimation results. Based on the observations in Table. \ref{tab:no_border} and the semi-major axis $a$, we can approximate the bounded range of the parameter as }
\bqn\lb{gggeq}
 \dot{f}_{\rm PV} &\lesssim& 10^{8} \rm m.
\eqn

At this gravity, the most relaxed constraints on the parity-violating parameters given by all current gravitational experiments come from the solar system experiments, and the result \red{we give here from the PSR J1141-6545 system is much weaker than those from the solar system $ \dot{f}_{\rm PV}\lesssim 10^{4} \rm m$ \cite{Qiao:2021fwi,Qiao:2023hlr}, for instance, as shown in Ref.~\cite{Qiao:2021fwi,Qiao:2023hlr}, both the measurements of the periastron precession of LAGEOS satellites and the spin precession of the gyroscopies in Gravity Probe B give the constraint as $ \dot{f}_{\rm PV}\lesssim 10^{4} \rm m$.}. Even though we are considering a system of two compact bodies, the contribution of the parity-violating terms to $\dot{x}_{\rm SO}$ is suppressed by the eccentricity $e$, when it is clear that the contribution produced by the parity-violating terms in the observations is very small. 

\subsection{\red{Parity-violationg }correction of boundary terms}

Previously, our analysis of the effect of parity violation terms on the evolution of the system's orbital plane was based only on solutions derived from equations describing the external fields of the source. The final calculations show that the eccentricity of the system orbit suppresses the contribution of these terms. However, since the field equations are considered to have continuity at the field-source boundary, it is clear that the solutions to these equations must contain an additional oscillatory component. Homogeneous solutions in CS gravity are given in Ref. \cite{Smith:2007jm}, which contain oscillatory terms that ensure that the gravitational field is continuous at the boundary of the source. These oscillatory terms produce contributions that distinguish them from external solutions in the observation of dynamical systems. Both solar system studies and studies of pericentre precession in pulsar systems have demonstrated that there is a corrective influence of these components on the evolution of the system's orbit that is not suppressed by eccentricity \cite{Smith:2007jm,Ali-Haimoud:2011wpu}. These findings therefore enable experimental observations to impose tighter constraints on the relevant theoretical parameters. We must therefore consider the effect of these additional oscillatory terms in the evolution of the orbital plane of the system, namely, the following terms in 
 \cite{Smith:2007jm} will be present in $\delta g_{0i}$ of Eq. (\ref{cs1})
\bqn
\delta \bl{g} = -\frac{15 G  }{2m_{\rm cs} R^3} \left[ C_1 \bl{S} + C_2 \bl{n} \times \bl{S} + C_3 \bl{n} \times (\bl{n} \times \bl{S}) \right]\nb\\
\lb{eqdcs}
\eqn
with
\bqn
C_1 &=&  \f{2R}{r}J_2(m_{\rm cs}R)Y_1(m_{\rm cs}r),\nb\\
C_2 &=& m_{\rm cs}R J_2(m_{\rm cs}R)Y_1(m_{\rm cs}r),\nb\\
C_3 &=& m_{\rm cs}RJ_2(m_{\rm cs}R)Y_2(m_{\rm cs}r),
\eqn
where the  spin angular momentum $\bl{S}=I\bl{\omega}$, with $I = 2mR^2/5$ is the moment of inertia and $m={4\pi R^3\rho}/{3}$ is mass of the sphere. $J_i$ and $Y_i$ are spherical Bessel functions of first and second kind. The quantity $m_{\rm cs}=1/\dot{f}_{\rm PV}$ represents the parity-violating parameter of the gravity, and here we adopt the notation from the literature to facilitate comparison with the results in Ref. \cite{Smith:2007jm}.

Next, we focus on investigating whether these oscillatory terms significantly influence the orbital evolution of the system, potentially yielding stronger constraints on the theoretical parameters through experimental observations. To achieve this, we calculate the corrections to the relative acceleration arising from the presence of these correction terms, following the previous procedure. We clearly observe that the metric in Eq. (\ref{eqdcs}) includes spherical Bessel functions associated with the parameter $m_{\rm cs}$. Here we introduce an approximation by retaining only the lowest-order contribution to the metric with respect to the parameter when calculating the corrections to the relative acceleration. Below, we directly present the calculated corrections as

\begin{widetext}
\bqn\lb{cscc}
\Delta a^i &=&-\f{1}{c^2} \f{15 G }{2m_{\rm cs}R^3_2}\sin(m_{\rm cs}R_2)\Bigg\{\f{m_1}{m}v^k \Bigg[ {\f{\sin(m_{\rm cs}r)}{r}{\omega}^i_{2} n^k}
 -{\f{\cos(m_{\rm cs}r)}{r} n^k\epsilon^{ijm}n^j{\omega}^m_2}
 -{\f{\sin(m_{\rm cs}r)}{r}n^kn^i(n^j{\omega}^j_2)}  \Bigg]\nb\\ 
&& +\f{m_2}{m}v^k \Bigg[{\f{\sin(m_{\rm cs}r)}{r} n^k{\omega^i}_2 }-{\f{\cos(m_{\rm cs}r)}{r}n^k\epsilon^{ijm}n^j{\omega}^m_2 }
-{ \f{\sin(m_{\rm cs}r)}{r}n^kn^in^j{\omega}^j_2}
 \Bigg]\nb\\
&& - \f{m_2}{m}v^k  \Bigg[{\f{\sin(m_{\rm cs}r)}{r} n^i{\omega^k}_2 } 
-{\f{\cos(m_{\rm cs}r)}{r}n^i\epsilon^{kjm}n^j{\omega}^m_2 }
-{ \f{\sin(m_{\rm cs}r)}{r}n^in^kn^j{\omega}^j_2}
 \Bigg]  \Bigg\}\nb\\
&&-\f{1}{c^2} \f{15 G\rho_1 }{2m_{\rm cs}R^3_1}{\sin(m_{\rm cs}R_1)}\Bigg\{\f{m_2}{m}v^k \Bigg[ {-\f{\sin(m_{\rm cs}r)}{r}{\omega}^i_{1} n^k} 
  -{\f{\cos(m_{\rm cs}r)}{r} n^k\epsilon^{ijm}n^j{\omega}^m_1}
 +{\f{\sin(m_{\rm cs}r)}{r}n^kn^i(n^j{\omega}^j_1)}  \Bigg] \nb\\
&& + \f{m_1}{m}v^k  \Bigg[{-\f{\sin(m_{\rm cs}r)}{r} n^k{\omega^i}_1 }-{\f{\cos(m_{\rm cs}r)}{r}n^k\epsilon^{ijm}n^j{\omega}^m_1 } +{ \f{\sin(m_{\rm cs}r)}{r}n^kn^in^j{\omega}^j_1}
 \Bigg]\nb\\
&& -\f{m_1}{m}v^k \Bigg[-{\f{\sin(m_{\rm cs}r)}{r} n^i{\omega^k}_1 }
 -{\f{\cos(m_{\rm cs}r)}{r}n^i\epsilon^{kjm}n^j{\omega}^m_1 }
+{ \f{\sin(m_{\rm cs}r)}{r}n^in^kn^j{\omega}^j_1}
 \Bigg]  \Bigg\}.
\eqn
\end{widetext}
In the equations above, $R_1$ and $R_2$ denote the radii of the pulsar and the white dwarf companion, respectively, and $m_1$ and $m_2$ represent their masses, and $m=m_1+m_2$ is the total mass of the system. 
In the above expression, we observe that while each term of the corrected acceleration is of order $1/r$, it also depends on the radius of the object as $1/R^3_i$(where $i=1,~2$). This is explicitly different from Eq. (\ref{GR1}) and Eq. (\ref{cs3}), where both equations are of order $1/r^3$ at least. The quantitative difference between the 
$1/R^3_i$ and $1/r^3$ magnitudes reflects the distinction between the contributions of Eq. (\ref{cscc}) and Eq. (\ref{cs3}) to the orbital evolution of the system.

To evaluate the contribution generated by the above correction terms, we use the orbital inclination element Eq. (\ref{w}) and average it by integrating over the interval $[0, 2\pi]$ of one orbital period. Note that in the integral treatment, to ensure the integrals are cumulative and to obtain an analytical expression, we can use the approximation $r(f)\approx a(1-e\cos f)$. This simplification allows us to proceed with the calculations while maintaining tractability and providing meaningful results.
After tedious calculations, we directly give the result as
\bqn\lb{ddmm}
\Delta \f{d\iota}{d t} &=&\f{1}{c^2}\f{1}{ a^{2}}\f{ J_1(m_{\rm cs}ae)}{m^2_{\rm cs}} \cos\omega \Bigg\{\f{15 G}{2R^3_2}{\sin(m_{\rm cs}R_2)}  \bl{S}_{2} \cdot \hat{\bl{h}}  \nb\\
&&-\f{15 G}{2R^3_1}{\sin(m_{\rm cs}R_1)}  \bl{S}_{1} \cdot \hat{\bl{h}}
\Bigg\}\sin(m_{\rm cs}a).
\eqn
From the above equation, it is evident that the contribution of the boundary terms arising from the parity-violating effect is perpendicular to the orbital plane, aligning with the direction specified in Eq. (\ref{dm2}). \red{However, unlike Eq. (\ref{dm2}), where the contribution depends linearly on the theoretical parameter $\dot{f}_{\rm PV}$, the contribution here is quadratically dependent on the theoretical parameters $1/m^2_{\rm cs}=\dot{f}^2_{\rm PV}$}. This distinction underscores the unique nature of the parity-violating boundary terms' influence on the system dynamics compared to the effects described in Eq. (\ref{ddmm}).

We focus primarily on the effect of the white dwarf companion on the evolution of the system's orbital plane, and therefore retain only the contribution of this object. Using Eq. (\ref{dm1}), we derive the contribution of the boundary terms to the temporal evolution of the projected semi-major axis as
\bqn\lb{x122}
\Delta\dot{x}_{\rm LT} &=& -x   \f{1}{c^2}\f{S_{2} }{ a^{2}}\f{15 G}{2R^3_2}\f{ J_1[m_{\rm cs}ae]}{m^2_{\rm cs}} \nb\\
&&\times \cos\omega {\sin(m_{\rm cs}R_2)}   
\sin(m_{\rm cs}a) \cot\iota \cos \delta_c.\nb\\
\eqn
This expression represents the contribution to the observable $\dot{x}$ over an orbital period, incorporating the parity-violating boundary terms. From this expression, it is evident that the result exhibits oscillatory behavior with respect to the theoretical parameter. To combine observations and derive constraints on these parameter, we perform further processing. Compared to the results of general relativity, we find that
\bqn
\f{\Delta\dot{x}_{\rm LT}}{\dot{x}_{GR}} &=& a(1-e^2)^{3/2}\f{15 }{2R^3_2}{\sin(m_{\rm cs}R_2)}\f{ J_1(m_{\rm cs}ae)}{m^2_{\rm cs}} \cos\omega  \nb\\
&&\f{\cot\delta_c}{\sin \Phi_c} \sin(m_{\rm cs}a)\f{2m_2}{4m_2+3m_1}. \label{cpd1}
\eqn

\begin{figure}[htbp]
    \centering
    \includegraphics[width=0.45\textwidth]{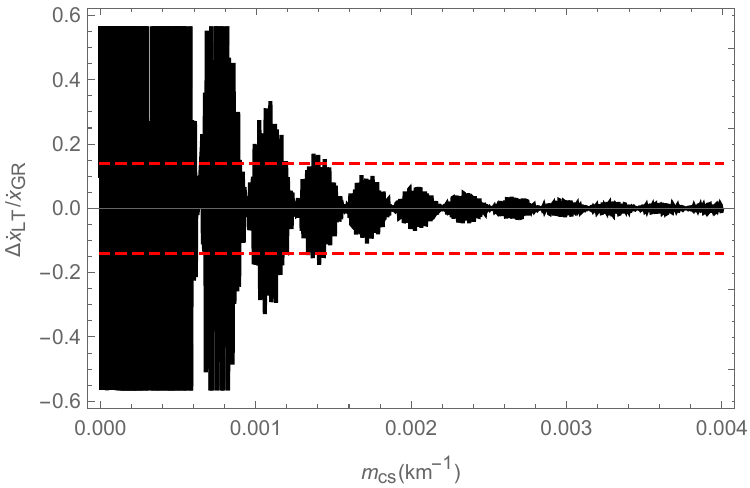} 
    \caption{The oscillating black solid line represents the variation of the $\Delta\dot{x}_{\rm LT}/\dot{x}_{GR}$ ratio with respect to the parameter, where $R_2=2.1\times 10^3~\rm km$. The 13.9\% verifcation of the observations (red dashed line) leads to \red{a limit of $m_{\rm cs}\gtrsim 0.14\times 10^{-2} \rm  km^{-1}$  for the parameter in the PSR J1141-6545}.} 
    \label{Fg1} 
\end{figure}
\begin{figure}[htbp]
    \centering
    \includegraphics[width=0.45\textwidth]{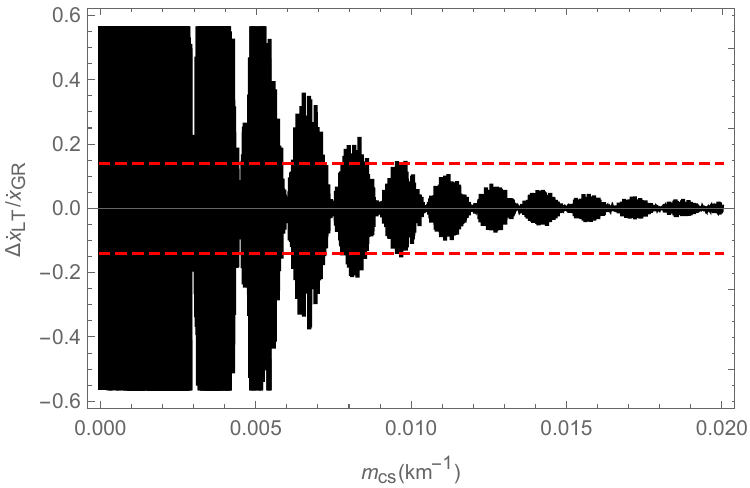} 
    \caption{The oscillating black solid line represents the variation of the $\Delta\dot{x}_{\rm LT}/\dot{x}_{GR}$ ratio with respect to the parameter, where $R_2=10^4~\rm km$. The 13.9\% verifcation of the observations (red dashed line) leads to \red{a limit of $m_{\rm cs}\gtrsim 0.95\times 10^{-2} \rm  km^{-1}$  for the parameter in the PSR J1141-6545}.} 
    \label{Fg2} 
\end{figure}
\begin{figure}[htbp]
    \centering
    \includegraphics[width=0.45\textwidth]{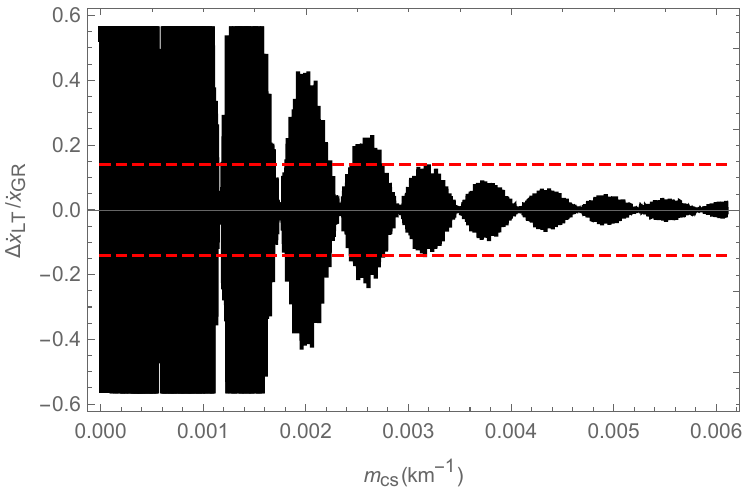} 
    \caption{The oscillating black solid line represents the variation of the $\Delta\dot{x}_{\rm LT}/\dot{x}_{GR}$ ratio with respect to the parameter, where $R_2=5.4\times 10^3~\rm km$. The 13.9\% verifcation of the observations (red dashed line) leads to \red{a limit of $m_{\rm cs}\gtrsim 0.32\times 10^{-2} \rm  km^{-1}$  for the parameter in the PSR J1141-6545}.} 
    \label{Fg3} 
\end{figure}
The observations $\dot{x}_{\rm obs} = (1.7 \pm 0.3)\times 10^{-13} \rm ss^{-1}$ in PSR J1141-6545, as listed in Table \ref{tab:no_border}, indicate that the uncertainty arising from spin-orbit interactions is approximately $13.9\%$ in the observed data. Consequently, we can constrain the theoretical parameters graphically by requiring the ratio of Eq. (\ref{cpd1}) to be less than $13.9\%$. In Eq. (\ref{cpd1}), two key parameters are involved: the white dwarf mass $M_2$ and radius $R_2$. It should be noted that while the mass parameter has been precisely determined through high-precision observations, the radius parameter remains subject to uncertainty. From a rigorous theoretical perspective, the most comprehensive approach would derive the mass-radius relation from the white dwarf's equation of state. However, to meet the practical requirements of this study, we adopt an alternative methodology that balances efficiency with reliability. By considering the currently observed reasonable range of white dwarf radii ($R_2 \in [2.1~\rm km\cite{Caiazzo:2021xkk}, 10~\rm km]$), we calculate the corresponding parameter constraint spaces for both the lower and upper radius limits. Through comparative analysis, we select the most inclusive constraint condition as the final allowable range for the theoretical parameters. Fig. \ref{Fg1} through Fig. \ref{Fg3} systematically present the variation of Eq. (\ref{cpd1}) with theoretical parameters under different white dwarf radius assumptions, each providing corresponding parameter constraints:  Fig. \ref{Fg1} adopts the lower-bound radius $R = 2.1\times 10^3~\rm km$ (most compact configuration), constraining the parameter to ${1}/{m_{\rm cs}} = \dot{f}_{\rm PV}\lesssim 1.0\times 10^5~ \rm m$. Fig. \ref{Fg2} assumes the upper-bound radius $R = 10\times 10^3~\rm  km$ (maximally extended case), yielding a relaxed constraint of ${1}/{m_{\rm cs}} = \dot{f}_{\rm PV}\lesssim 7.1\times 10^5~ \rm m$. Fig. \ref{Fg3} implements the canonical radius R = 5.4 × 10³ km from Ref. \cite{VenkatramanKrishnan:2020pbi}, resulting in $1/m_{\rm cs} = \dot{f}_{\rm PV} \lesssim 3.3\times10^5~ {\rm m}$.

Through comparative analysis of these three scenarios, we establish the final constraint on the theoretical parameter as $1/m_{\rm cs} = \dot{f}_{\rm PV} \lesssim 10^6~ {\rm m}$. It is evident that the bound on the parameter $\dot{f}_{\rm PV}$ is two orders of magnitude tighter than the one obtained in Eq. (\ref{gggeq}). In the above three cases, although we did not consider the specific equation of state of white dwarfs, by taking into account the most extreme boundary values of the current white dwarf radii, we provided constraints corresponding to the parameters respectively. From the results, when the white dwarf radii take the maximum and minimum values respectively, the constraints on the corresponding theoretical parameters only differ within an order of magnitude. This fully indicates that considering the mass-radius relationship given by the equation of state of white dwarfs has a very small impact on the constraints of our final parameter.

\begin{figure}[htbp]
    \centering
    \includegraphics[width=0.45\textwidth]{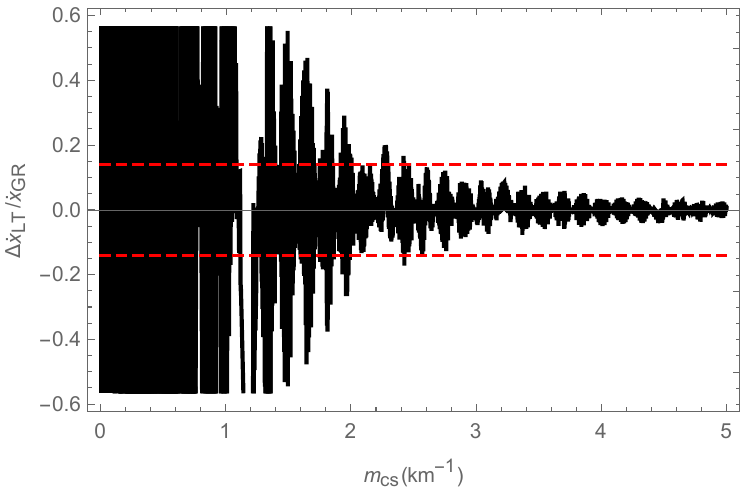} 
    \caption{The oscillating black solid line represents the variation of the $\Delta\dot{x}_{\rm LT}/\dot{x}_{GR}$ ratio with respect to the parameter, where the radius of pulsars $R_1=20~\rm km$. The 13.9\% verification of the observations (red dashed line) leads to {a limit of $m_{\rm cs}\gtrsim 2.5~ \rm  km^{-1}$  for the parameter in the PSR J1141-6545}.} 
    \label{Fg4} 
\end{figure}

Furthermore, as demonstrated in our earlier analysis, the pulsar's contribution currently falls below the detection threshold of present measurement uncertainties. We anticipate that future advancements in observational precision may render these contributions detectable. To systematically evaluate the pulsar's capacity for constraining theoretical parameters within this system, we maintain a current measurement precision of $13.9\%$ as a conservative baseline. By incorporating specific pulsar parameters into Eq. (\ref{cpd1}) with an upper radius bound of $R_1= 20 ~\rm km$ based on the pulsar, we plot the corresponding parameter variations presented in Fig. \ref{Fg4}. We can derive that the constraint on the parameter is ${1}/{m_{\rm cs}} = \dot{f}_{\rm PV}\lesssim 0.4~ \rm km$, which represents an improvement of four orders of magnitude compared to the constraints obtained from white dwarfs.

In this gravitational framework, the parity-violating terms generated simultaneously outside the field source and at the boundary exert an influence on the evolution of the system's orbital plane. Using experimental observations of white dwarfs in binary systems, we have respectively derived the constraints on the theoretical parameters arising from these two types of corrections. By comparing the two constraint ranges, we obtain the tightest constraint on the theoretical parameter from the Lense-Thirring effect of white dwarfs:  $ \dot{f}_{\rm PV}\lesssim 10^6~ \rm m$, which corresponds to an energy scale $M_{\rm PV}:= 1/\dot{f}_{\rm PV} \gtrsim 10^{-22} \rm GeV$. This constraint range differs by only two orders of magnitude from that derived from the solar system ($ \dot{f}_{\rm PV}\lesssim 10^{4} \rm m$) \cite{Qiao:2021fwi,Qiao:2023hlr}. We further predict the contribution of observing the pulsar in this system in the future. Using pulsar observations, the constraint on the theoretical parameter can be further improved to $ \dot{f}_{\rm PV}\lesssim 10^2~ \rm m$, which corresponds to an energy scale $M_{\rm PV}:= 1/\dot{f}_{\rm PV} \gtrsim 10^{-18} \rm GeV$. We find that this constraint is at the same level as the constraint on the parity-violating parameter derived from the pericenter precession in pulsar binary systems \cite{Ali-Haimoud:2011wpu}. In Table \ref{tab2}, we present the constraints on the energy scales associated with the parameters in ghost-free parity-violating and CS gravity, derived from a range of experimental observations. The constraints reported in this section apply to both CS and ghost-free parity-violating gravity. For ghost-free parity-violating gravity, the constraint $ \dot{f}_{\rm PV}\lesssim 10^6~ \rm m$, while less stringent than that derived from GW events, is close to the constraints inferred from solar system measurements. 

\begin{table}[htbp]
    \centering
    \caption{Constraints on the parity-violating energy scale $M_{\rm PV}$ form different experiments}
    \label{tab2}
    \begin{tabular}{cccc}
        \hline\hline 
       Theories  & Solar system  & Compact binary & GWs \\ \hline 
       \makecell[ct]{ \rule{0pt}{10pt} CS gravity(GeV)}  &   \makecell[ct]{ \rule{0pt}{10pt} $10^{-22}$ \cite{Smith:2007jm}} &      \makecell[ct]{\rule{0pt}{10pt}$ 10^{-18}$ \cite{Ali-Haimoud:2011wpu}}  & \makecell[ct]{\rule{0pt}{10pt} $10^{-22}$ \cite{Wang:2020cub}} \\  \hline
       \makecell{Ghost-free\\ parity-violating\\ gravity(GeV)}  &    $10^{-20}$ \cite{Qiao:2021fwi,Qiao:2023hlr} &       \makecell{$10^{-22}$(WD)\\$10^{-18}$(pulsar)}   &$10^{-2}$\cite{Wang:2020cub}  \\
       \hline \hline
    \end{tabular}
\end{table}

\section{conclusions}
\renewcommand{\theequation}{5.\arabic{equation}} \setcounter{equation}{0}
\lb{sec5}

In this work, we extend previous studies by employing the observable progression of the orbital plane inclination to test for gravitational parity violation in the PSR J1141-6545 system. The misalignment between the white dwarf's spin axis and the system's total angular momentum induces a precession effect in the orbital plane inclination, which can be detected through the temporal evolution of the projected semi-major axis. We adopt a parity-violating gravitational framework, where the spacetime metric incorporates parity-violating terms associated with both the exterior and boundary of the field source, in contrast to GR. 
Within this framework, we calculate the relative acceleration of the system and find that the corrected relative acceleration arising from the two-part parity-violating terms exhibit an inverse proportionality to $r$ of different orders. We further compute the corrections to the orbital inclination precession rates induced by the two-part parity-violating terms, which demonstrate significant deviations from the GR prediction. Specifically, the parity-violating contribution to the orbital inclination precession rates depends on the projection of the spin vector along the orbital angular momentum direction, whereas in GR, this contribution arises from the projection of the spin vector within the orbital plane. The parity-violating terms generate corrections to the orbital inclination precession rates that are perpendicular to the GR, highlighting a fundamental distinction between the parity-violating framework and GR.

Furthermore, the two orbital inclination precession rates corresponding to the parity-violating corrections exhibit distinct dependencies on the orders of the theoretical parameters.  The correction associated with the gravitational field outside the source is linear in the theoretical parameter and coupled to the eccentricity $e$, while the correction describing the boundary terms is quadratic in the theoretical parameter. These differences reflect the distinct contributions of the two parity-violating components to the orbital inclination precession effect. By comparing the orbital inclination precession rates of the two parity-violating corrections with those of GR, and incorporating the uncertainties in the observational data, we derive the  constraints on the theoretical parameters as 
$ \dot{f}_{\rm PV}\lesssim 10^6~ \rm m$. The constraints on the energy scales associated with ghost-free parity-violating and CS gravity, derived from a range of experimental observations, reveal distinct levels of sensitivity for the theories. 
However, for ghost-free parity-violating gravity, the constraint $ \dot{f}_{\rm PV}\lesssim 10^6~\rm m$, while less stringent than that derived from GW events, is close to the constraint inferred from solar system measurements. \red{ We also predict that when the pulsar is observed in the future, the constraint on the theoretical parameter can be further improved to $ \dot{f}_{\rm PV}\lesssim 10^2~\rm m$ using pulsar data, which is consistent with the constraint derived from the periastron precession of pulsars \cite{Ali-Haimoud:2011wpu}.
This indicates that compact binary pulsar observations provide a more sensitive probe for testing parity-violating gravity compared to traditional solar system tests.}

\red{In this work, we focus on the parity-violation-induced corrections to the spin-orbit interaction,
which are then tested against current observational constraints. Given that the pulsar's contribution currently falls below the measurement uncertainty threshold, we may provisionally neglect the effects arising from its spin dynamics. Continued monitoring of the binary system PSR J1141-6545 will likely yield improved observational precision. As the measurement accuracy approaches the magnitude of the pulsar's contribution, this previously negligible effect will become statistically significant. At such time, we propose to incorporate this parameter into our result, enabling a refined constraint of the theoretical parameter. Additionally, in the study of the binary system PSR J1141-6545, the contribution from the mass quadrupole moment becomes increasingly significant as the white
dwarf’s spin period decreases. Therefore, incorporating parity-violation effects on the mass quadrupole
moment could provide an additional avenue for theoretical validation, leveraging this mechanism to further
constrain model parameters.}

\section*{Acknowledgements}

This work is supported by the National Natural Science Foundation of China under Grants No. 12275238, No. 11975203, No. 11675143, the National Key Research and Development Program of China under Grant No. 2020YFC2201503, the Zhejiang Provincial Natural Science Foundation of China under Grants No. LR21A050001 and No. LY20A050002, and the Fundamental Research Funds for the Provincial Universities of Zhejiang in China under Grant No. RF-A2019015. Wen Zhao is supported by  the National Natural Science Foundation of China  under Grants No. 12325301 and 12273035.

\section*{Appendix: PPN potentials}
\renewcommand{\theequation}{A.\arabic{equation}} \setcounter{equation}{0}

In this Appendix, we present the explicit expressions for the PPN potentials used to parameterize the metric in Eqs. (\ref{meq1}). These potentials are given as follows \cite{will2018th}:
\bqn \label{App}
U & \equiv & \int \frac{\rho\left(\mathbf{x}^{\prime}, t\right)}{\left|\mathbf{x}-\mathbf{x}^{\prime}\right|} d^{3} x^{\prime}, \lb{SPU}\\
\Psi&=&2\Phi_1 - \Phi_2 +\Phi_3 +4\Phi_4-\f{1}{2}\Phi_5-\f{1}{2}\Phi_6,\\
\Phi_{1} & \equiv & \int \frac{\rho^{\prime} v^{\prime 2}}{\left|\mathbf{x}-\mathbf{x}^{\prime}\right|} d^{3} x^{\prime} ,\\
\Phi_{2} & \equiv & \int \frac{\rho^{\prime} U^{\prime}}{\left|\mathbf{x}-\mathbf{x}^{\prime}\right|} d^{3} x^{\prime}, \\
\Phi_{3} & \equiv & \int \frac{\rho^{\prime} \Pi^{\prime}}{\left|\mathbf{x}-\mathbf{x}^{\prime}\right|} d^{3} x^{\prime}, \\
\Phi_{4} & \equiv & \int \frac{p^{\prime}}{\left|\mathbf{x}-\mathbf{x}^{\prime}\right|} d^{3} x^{\prime} \\
\Phi_{5} & \equiv & \int \frac{\rho^{\prime}\left[\mathbf{v}^{\prime} \cdot\left(\mathbf{x}-\mathbf{x}^{\prime}\right)\right]^{2}}{\left|\mathbf{x}-\mathbf{x}^{\prime}\right|^{3}} d^{3} x^{\prime}, \\
\label{App1} \Phi_{6} & \equiv &\int \rho^{\prime} \mathbf{\nabla}^{\prime} U^{\prime} \cdot   \frac{\left(\mathbf{x}-\mathbf{x}^{\prime}\right)  }{\left|\mathbf{x}-\mathbf{x}^{\prime}\right|} d^{3} x^{\prime}.
\eqn

\appendix

\end{document}